\documentclass[conference]{IEEEtran}
\IEEEoverridecommandlockouts
\usepackage{cite}
\usepackage{amsmath,amssymb,amsfonts}
\usepackage{algorithmic}
\usepackage{graphicx}
\usepackage{textcomp}
\usepackage{xcolor}
\def\BibTeX{{\rm B\kern-.05em{\sc i\kern-.025em b}\kern-.08em
    T\kern-.1667em\lower.7ex\hbox{E}\kern-.125emX}}
\begin{document}

\title{A Stochastic Mean-CVaR Framework for BESS Multi-Market Bidding Strategies}

\author{\IEEEauthorblockN{Younes Zahraoui}
\IEEEauthorblockA{\textit{LIST, Luxembourg} \\
https://orcid.org/0000-0002-0505-3553}
\and
\IEEEauthorblockN{Jun CAO}
\IEEEauthorblockA{\textit{LIST, Luxembourg} \\
https://orcid.org/0000-0001-5099-9914}
\and
\IEEEauthorblockN{Samir Kouro}
\IEEEauthorblockA{\textit{LIST, Luxembourg} \\
https://orcid.org/0000-0002-1690-4624}
\and
\IEEEauthorblockN{Pedro Rodriguez Cortes}
\IEEEauthorblockA{\textit{LIST, University of Luxembourg} \\
https://orcid.org/0000-0002-1865-0461}

}

\maketitle

\begin{abstract} 
Battery Energy Storage Systems (BESS) operators face significant challenges when participating in multiple electricity markets due to the complex coupling of price volatility and stochastic reserve activation. Traditional deterministic dispatch models  neglect the "tail risks" associated with extreme market realizations, potentially leading to technical infeasibility or severe economic losses. This paper proposes a risk-aware stochastic optimization framework for the co-optimization of BESS participation in the Day-Ahead (DA) energy market and the manual Frequency Restoration Reserve (mFRR) market. The model explicitly captures multi-dimensional uncertainties by utilizing non-parametric Kernel Density Estimation (KDE) to generate joint price scenarios that preserve the empirical characteristics of European balancing markets. A two-stage stochastic programming approach is employed.To manage the financial exposure to high-impact price events, the Conditional Value-at-Risk (CVaR) metric is integrated into a Mean-CVaR objective function. This allows decision-makers to tune their risk-aversion levels and identify an efficient frontier between profitability and robustness. Simulation results demonstrate that the proposed joint CVaR approach significantly enhances revenue stability compared to deterministic benchmarks.

\end{abstract}


\section{Introduction}
The rapid expansion of variable renewable energy sources has intensified the need for flexible and reliable assets capable of balancing supply and demand in increasingly decarbonized power systems [1]. Among the available flexibility options, Battery Energy Storage Systems (BESS) have emerged as a pivotal technology for enabling the green transition. By providing fast response services such as energy arbitrage, frequency regulation, and balancing reserves, BESS facilitate higher renewable penetration, enhance system stability, and reduce reliance on fossil-fuel-based peaking units [2].

Beyond their technical role, BESS are becoming central market participants in modern electricity market designs [3]. In Europe and globally, regulatory reforms and national energy strategies increasingly recognize storage as a key enabler of climate neutrality, energy security, and grid resilience. Policy frameworks such as the European Green incentives, the Clean Energy Package, and various national storage roadmaps aim to accelerate large-scale deployment of BESS technologies by clarifying market access, revenue stacking opportunities, and participation in ancillary service markets [4].

To effectively monetize this flexibility, BESS operators must navigate a complex landscape of sequential electricity markets [5]. In the European context, the power exchange is structured around three primary pillars: the Day-Ahead (DA) market, which serves as the main arena for scheduling bulk energy production and consumption; the Intraday (ID) market, utilized for continuous adjustment of positions closer to real-time; and the Ancillary Services (AS) market, which is critical for maintaining physical grid stability [6].

\begin{figure}[t]
\centerline{\includegraphics[width=0.5\textwidth, height=5cm]{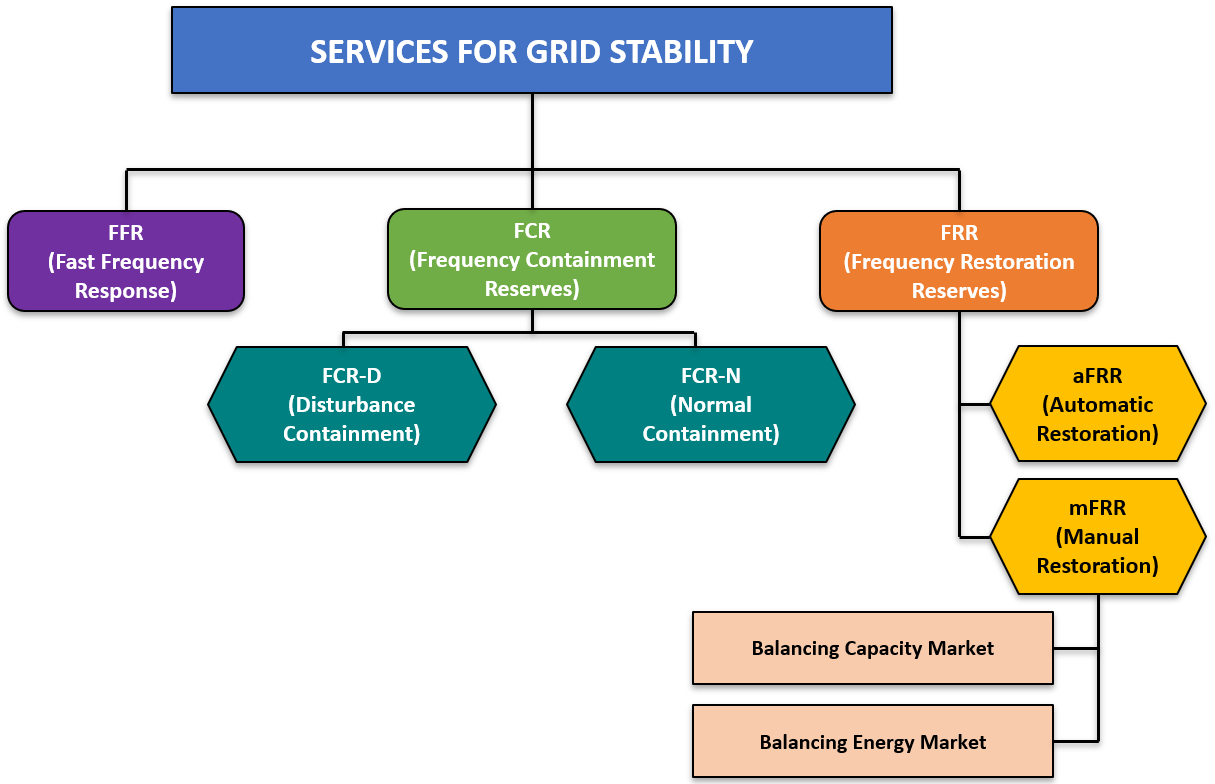}}
\caption{Hierarchical Structure of European Balancing and Frequency Reserves.}
\label{fig}
\end{figure}

The AS market is hierarchically organized into distinct reserve products designed to address frequency deviations at different timescales as illustrated in Fig 1. The first line of defense is the Frequency Containment Reserve (FCR). Activated automatically within seconds of a disturbance, FCR stabilizes the system frequency across the entire synchronous area, requiring providers to ramp fully within 30 seconds. While FCR arrests the initial frequency deviation, it stabilizes the system at a quasi-steady state rather than restoring the nominal 50/60 Hz frequency. The subsequent restoration task is performed by Frequency Restoration Reserves (FRR), which are subdivided into automatic (aFRR) and manual (mFRR) activation products. These reserves are deployed regionally within specific Load Frequency Control (LFC) areas to compensate for power imbalances and restore cross-border flows to their scheduled values. Finally, Replacement Reserves (RR) may be called upon to free up FRR capacity for future events. This tiered structure offers unique revenue stacking opportunities for fast-acting storage assets, particularly in restoration reserves where BESS can capitalize on both capacity availability and energy activation [7].

The literature has extensively documented the revenue potential for the flexible resources through the provision of AS. For example, In [4] a quantitative framework is introduced to analyze the coupling between imbalance settlement rules and TSO activation strategies in multi-product balancing markets. The study demonstrates that while pricing based solely on mFRR discourages strategic self-dispatching, settlement schemes averaging mFRR and aFRR prices can actually reduce total system costs when the TSO employs suboptimal reserve activation protocols. Conversely, more complex pricing mechanisms often distort signals and induce inefficient reactive balancing. In [6] the authors bridge the gap between economic scheduling and the physical limitations of distributed energy resources by proposing a stochastic mixed-integer linear programming (MILP) model. The framework co-optimizes aggregator portfolios across Day-Ahead and FCR markets while strictly enforcing dynamic grid-code constraints. The study reveals that explicitly modeling actual reserve activation and State of Energy (SOE) recovery drives storage technologies to adopt more risk-averse strategies, shifting their capacity allocation away from reserve provision and toward the energy market. In [8], a comprehensive mixed-integer linear programming model is developed for the cross-border clearing of mFRR under the European MARI framework. The authors integrate both self-scheduling and central-scheduling market designs, introducing a novel pre-processing step that converts diverse bids into standard balancing products. By explicitly incorporating complex operational features—such as varied order types, TSO tolerance bands, and cross-zonal capacity constraints—their simulations across twenty-five European control areas demonstrate that joint clearing significantly reduces overall balancing costs and promotes regional price convergence through implicit imbalance netting. In [9] a mixed-integer linear programming model is developed to co-optimize electricity procurement and reserve provision for an electrolyzer. The authors analytically derive the opportunity cost of participating in frequency-supporting markets, specifically comparing FCR with energy-intensive mFRR. Using historical data from the Danish market, the study demonstrates that flexible bidding strategies and adaptable hydrogen demand structures can mitigate production disruptions, rendering mFRR particularly lucrative and capable of increasing operational profits by up to 47 $\%$.

To manage the extreme stochasticity and operational complexities of modern European balancing platforms, recent literature has increasingly shifted toward multi-stage stochastic frameworks. For instance, [1] investigates the optimal bidding strategy for a joint wind-battery asset. The research employs a three-stage stochastic model with CVaR to navigate the uncertainties inherent in day-ahead energy prices, aFRR capacity markets, and real-time energy activation via the PICASSO mechanism. The results emphasize that the financial viability of such pooled assets relies heavily on the underlying imbalance pricing structure, which dictates the value of using the BESS for internal wind error mitigation versus external market participation. In [10], the authors address the stochasticity of distributed energy resources by developing a chance-constrained optimization framework for an electric vehicle aggregator participating in the asymmetric FCR-D market. To satisfy strict transmission system operator reliability mandates without incurring the heavy computational burden of large sample-based methods, the study incorporates Extreme Value Theory (EVT). This analytical approach efficiently characterizes the tail behavior of EV flexibility, allowing the aggregator to construct highly reliable and computationally tractable bids that significantly reduce out-of-sample violation rates. [11], the authors investigate the computational challenges faced by EV aggregators participating in day-ahead and reserve markets, specifically focusing on the stringent minimum-volume bidding requirements. The study evaluates and adapts several state-of-the-art scheduling formulations—including stochastic, probabilistic, and virtual battery models—to manage these shared fleet constraints. Their comparative analysis reveals that while rigorous stochastic approaches yield the highest solution quality, the virtual battery method provides the optimal trade-off between economic performance, user demand satisfaction, and computational scalability. In [12] the authors propose a multi-stage stochastic optimization framework that integrates long-term investment planning with short-term bidding strategies for industrial flexible assets across spot and reserve markets. By evaluating the impacts of volatile electricity prices and rising carbon costs, the study demonstrates that anticipating reserve market participation fundamentally alters the optimal sizing of energy storage.

\section{contribution}

Despite recent advancements in multi-market scheduling, several critical gaps remain in the literature. Most existing dispatch models rely on deterministic approaches that neglect the stochastic nature of the prices, and the reserve activation. Furthermore, while stochastic frameworks have been proposed, they frequently assume standard parametric probability distributions that fail to capture the non-Gaussian, extreme volatility characteristic of European balancing markets. Finally, the specific co-optimization of BESS across DA and mFRR markets remains underexplored. To address existing gaps, this paper proposes a risk-aware stochastic framework for co-optimizing BESS in DA and mFRR markets. Our primary contributions are threefold:

- We develop a dispatch model coupling DA and mFRR capacity commitments with real-time mFRR energy activation uncertainty, overcoming the technical and economic vulnerabilities of traditional deterministic models.

- We employ non-parametric KDE to generate joint price scenarios. This accurately captures the extreme volatility, complex correlations, and non-Gaussian "fat-tailed" characteristics inherent in European balancing markets.

- We integrate the CVaR metric into the objective function. This provides decision-makers with a tunable tool to establish an efficient frontier between profitability and robustness, significantly enhancing revenue stability against extreme market events.

The remainder of this paper is organized as follows: Section III introduces the methodology framework for BESS participation in DA and AS markets. Section IV outlines the case study. The conclusions are presented in Section V.

\section{Methodology}
\subsection{Trading framework for renewable energy producers under multi- market coupling}

This study develops a stochastic optimization framework for the co-optimization of a BESS participating simultaneously in the DA energy market and the mFRR market. The BESS operator determines a unified operating schedule that maximizes economic performance while accounting for the intrinsic coupling between market uncertainty and physical operational constraints. The complexity of the problem arises from the two-stage structure of the mFRR market. In the first stage, the BESS commits reserve capacity and is remunerated for maintaining availability to provide upward ($r_{up}$) or downward ($r_{dn}$) regulation. In the second stage, the actual energy activation is a stochastic process driven by the TSO, where the resulting energy throughput depends on the expected activation rate ($\gamma$), which directly influences the State of Energy (SoE) dynamics of the BESS. Traditional dispatch models typically rely on average price forecasts, producing rigid schedules that only work under ideal conditions. If the market experiences sudden price shocks or extreme reserve calls, these deterministic approaches can cause severe financial losses or push the battery into technical failure, such as prematurely depleting its State of Charge (SoC). To address these limitations, a risk-aware stochastic optimization approach is adopted. By incorporating CVaR into the objective function, the proposed framework explicitly accounts for downside risk and enables a tunable trade-off between maximizing expected revenue and protecting the BESS operator against high-loss tail scenarios induced by price volatility and uncertain reserve deployment.

\subsection{Handling uncertainties in the prices}

Electricity market prices particularly in balancing markets like mFRR shows extreme volatility, non-Gaussian distributions, and complex temporal correlations. For a BESS, uncertainty is twofold: it involves both the clearing price risk and the activation volume risk. To capture this multi-dimensional uncertainty, we adopt a scenario-based stochastic framework. A finite set of scenarios $s \in \{1,\dots,S\}$ is constructed, where each scenario represents a joint realization of Day-Ahead ($\lambda_{DA,s,t}$), mFRR-up ($\lambda_{up,s,t}$), and mFRR-down ($\lambda_{dn,s,t}$) price trajectories. Unlike simple Monte Carlo sampling, these scenarios are weighted by probabilities $\pi_{s,t}$ derived from non-parametric Kernel Density Estimation (KDE). This approach ensures that the model respects the empirical characteristics of historical data such as the price "spread" between DA and mFRR without assuming a specific underlying distribution. The optimization is structured as a two-stage stochastic problem. The first-stage decisions comprising the DA energy commitments ($p_{ch,t}, p_{dis,t}$) and the reserved mFRR capacities ($r_{up,t}, r_{dn,t}$) to participate in mFRR market. These must be determined before the market clears and remain uniform across all scenarios to satisfy the non-anticipativity constraint. The second stage represents the scenario-dependent realization of profit or loss, which is contingent on the specific price paths and the expected activation rate ($\gamma$). Given that BESS profitability is highly sensitive to extreme price events (spikes and negative prices), risk is quantified using the Conditional Value-at-Risk (CVaR) metric. While standard variance-based risk measures treat upside and downside volatility equally, CVaR focuses exclusively on the "tail risk" the expected loss in the worst $(1-\alpha)\%$ of scenarios. By optimizing a Mean-CVaR objective function, the BESS operator can tune a risk-aversion parameter ($\beta$) to find an efficient frontier between aggressive profit-seeking and a robust defensive posture that protects the BESS from insolvency during periods of extreme market stress.

\subsection{Multi-market trading optimization model}

This study establishes a stochastic optimization model for BESS trading strategies across coupled markets, considering energy arbitrage in the Day-Ahead market and capacity provision in the upward and downward mFRR markets.

\subsubsection{Objective function}
The objective is to minimize the risk-adjusted loss of the BESS operator across all scenarios. Unlike benefit-only models, this formulation explicitly balances expected profit against the CVaR to ensure robustness against price volatility and activation uncertainty. The objective function is defined in Eq. (1):

\begin{equation}
    \min \quad (1-\beta) \cdot \mathbb{E}[L] + \beta \cdot \text{CVaR}_{\alpha}
\end{equation}

$\beta$ is the risk-aversion coefficient. The total expected loss $\mathbb{E}[L]$ is derived from the negative sum of revenues from the Day-Ahead market and the mFRR capacity markets, as described by Eqs. (2)–(4):

\begin{equation} 
\begin{aligned}
L_s = - \sum_{t \in T}  \Big[ &\lambda_{DA, s, t}(p_{dis, t} - p_{ch, t}) + \lambda_{up, s, t} \cdot r_{up, t} \\
 &+ \lambda_{dn, s, t} \cdot r_{dn, t} \Big] \cdot \Delta t
\end{aligned}
\end{equation}

\begin{equation}
\mathbb{E}[L] = \sum_{s \in S} \pi_s \cdot L_s 
\end{equation}

\begin{equation}
\text{Profit}_{s}= - L_s 
\end{equation}

where $\lambda_{DA, s, t}$ is the price value of the day-ahead at each scenario $s$, time $t$. $p_{dis, t}$ and $ p_{ch, t}$ represent the power dispatch (charging/discharging) for the BESS in the day-ahead. $\lambda_{up, s, t}$ and $\lambda_{dn, s, t}$ are the prices for the up and down regulation market respectively, at each scenario, $s$, and time $t$. $r_{up, t}$ and $r_{dn, t}$ are the power capacity for participating in the up, and down regulation, respectively. $\pi_s$ represents the Probability weight for each scenario $s$.

To reflect market rules where upward regulation is associated with discharging and downward with charging, and to prevent simultaneous opposing actions, we utilize a binary variable $\delta_{up}$ and $\delta_{dn}$.

\begin{equation}
    p_{dis, t} + r_{up, t} \leq P_{max} \cdot \delta_{up, t}
\end{equation}

\begin{equation}
    p_{ch, t} + r_{dn, t} \leq P_{max} \cdot \delta_{dn, t}
\end{equation}

\begin{equation}
    \delta_{up, t} + \delta_{dn, t} \leq 1, \quad 
\end{equation}

Where $\delta_{up, t}, \delta_{dn, t} \in \{0, 1\}$, The SoE transition incorporates the expected activation rate $\gamma$ to model the most likely energy throughput:

\begin{equation}
\begin{aligned}
    \text{SOE}_{t+1} = \text{SOE}_t + \Bigg[ &\eta \cdot (p_{ch, t} + \gamma_{dn} r_{dn, t}) \\
    &- \frac{p_{dis, t} + \gamma_{up} r_{up, t}}{\eta} \Bigg] \cdot \Delta t 
\end{aligned}
\end{equation}

Where $\gamma_{up}, \gamma_{dn}$ represent expected activation rates for reserved capacity.
$\eta$ is the efficiency of the BESS.

To ensure the BESS can physically fulfill 100$\%$ of its reserve commitment without violating energy limits, the following safety buffers are enforced:

\begin{equation}
    \text{SOE}_t - \left( \frac{p_{dis, t} + r_{up, t}}{\eta} \right) \cdot \Delta t \geq \text{SOE}_{min}
\end{equation}

\begin{equation}
    \text{SOE}_t + \left( \eta \cdot(p_{ch, t} + r_{dn, t}) \right) \cdot  \Delta t \leq \text{SOE}_{max}
\end{equation}

Market price uncertainties and stochastic reserve activations lead to discrepancies between planned and actual revenues. To address this, we integrate CVaR theory to ensure revenue stability.Value-at-Risk (VaR) represents the maximum potential loss at a confidence level $\alpha$. CVaR extends this by calculating the expected loss in the $(1-\alpha)\%$ of scenarios that exceed the VaR threshold:

\begin{equation}
    \text{CVaR}_{\alpha} = \zeta + \frac{1}{1 - \alpha} \sum_{s \in S} \pi_s \cdot u_s 
\end{equation}

Where $\alpha$ is the confidence level for risk assessment, $\zeta$ is the VaR threshold and $u_s$ is an auxiliary decision variable representing the loss for scenario $s$, defined by the following linear constraints:

\begin{equation}
    u_s \geq L_s - \zeta
\end{equation}

\begin{equation}
    u_s \geq 0
\end{equation}

By adjusting the risk-aversion parameter $\beta$ in Eq. (1), the operator can traverse the efficient frontier, balancing the drive for maximum profit against the necessity of limiting exposure to extreme market events.

\section{Case of study}

To validate the effectiveness of the proposed risk-aware co-optimization framework, a comprehensive case study is conducted using real-world market data. The study analyzes the optimal day-ahead bidding strategy of a grid-scale BESS participating simultaneously in the DA energy market and the mFRR market. The numerical analysis is based on historical DA market data obtained from the Luxembourg bidding zone. We utilise a non-parametric KDE technique to generate 100 joint price and activation scenarios based probability for each market. Figure 1 illustrates the DA electricity price trajectories under market uncertainty for 100 scenario. Figure 2 shows the KDE-Based probability density heatmap for DA market scenarios for each hour. 

The simulated asset is a utility-scale lithium-ion BESS. The power capacity, which dictates the maximum charging and discharging limits, is set to 0.15, and 0.95 $\%$ while the total energy capacity is 10 MWh, with maximum/minimum power limits of 1 MW. The efficiency is assumed to be 90 $\%$ for charging and discharging. The initial value of SOE is considered as 50 $\%$ of the maximum energy capacity. The Mean-CVaR optimization model is formulated as a MILP. Financial risk is quantified using a confidence level of 0.95, focusing the CVaR metric on the worst 5$\%$ of market realizations. To trace the efficient frontier, the risk-aversion weighting factor is systematically varied from 0.0 to 1.0. The optimization framework is implemented in Python and solved using the CVXpy.

\begin{figure}[htbp]
\centerline{\includegraphics[width=0.5\textwidth]{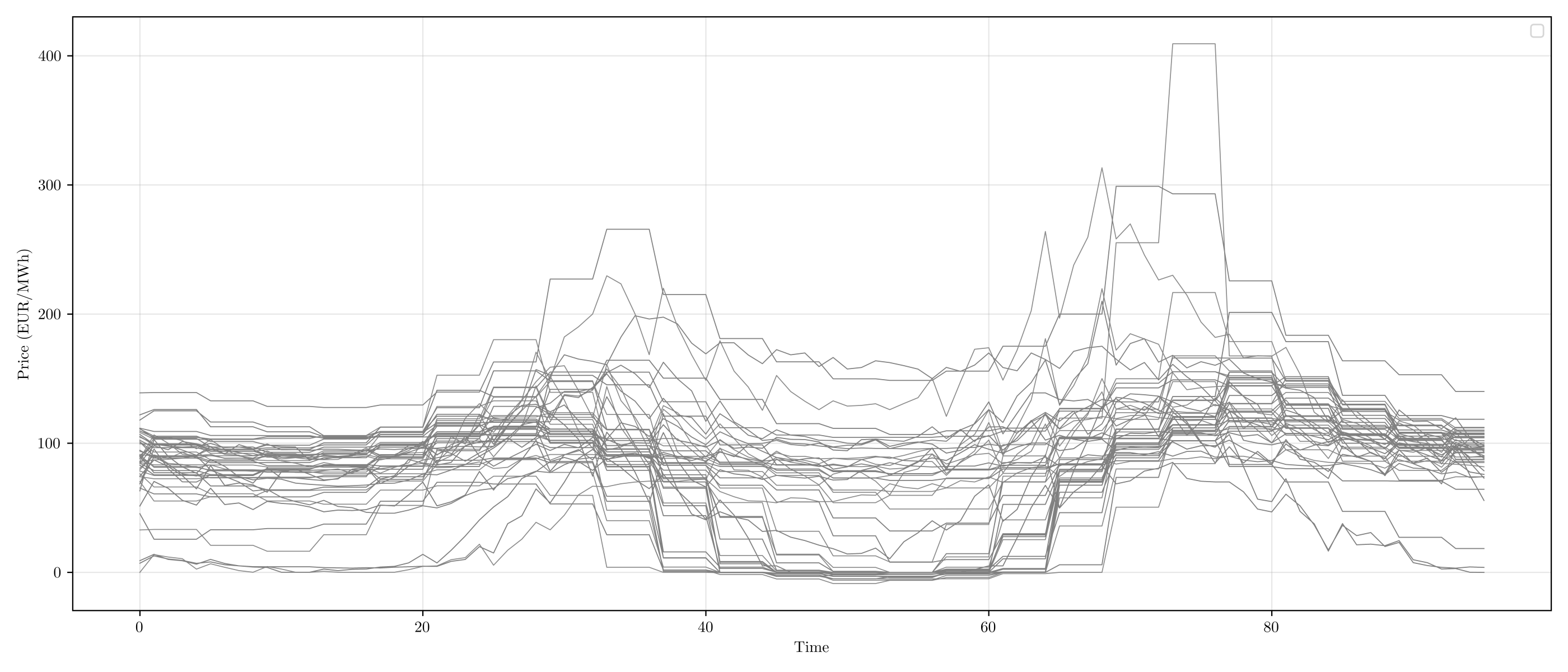}}
\caption{Day-Ahead electricity price trajectories under market uncertainty.}
\label{fig}
\end{figure}

\begin{figure}[htbp]
\centerline{\includegraphics[width=0.5\textwidth]{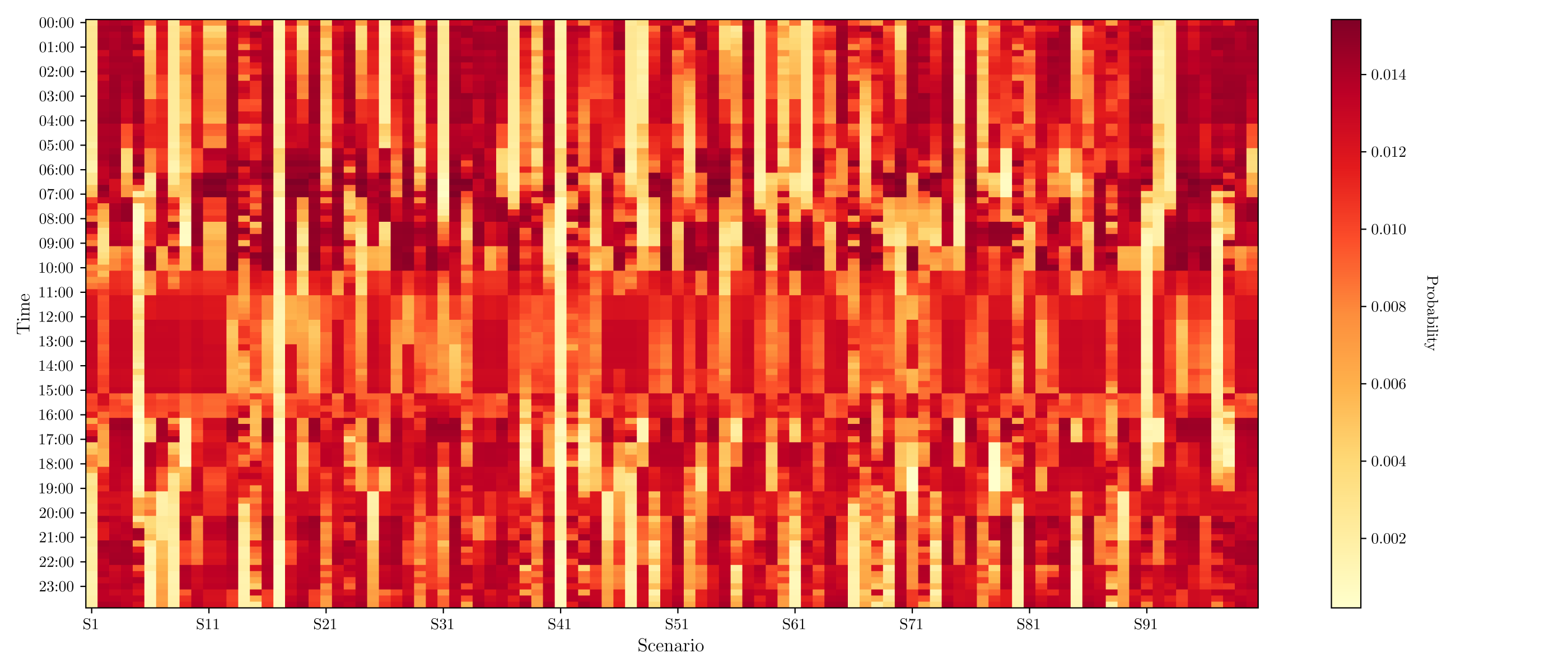}}
\caption{KDE-Based probability density heatmap for DA market scenarios.}
\label{fig}
\end{figure}

\begin{figure}[htbp]
\centerline{\includegraphics[width=0.5\textwidth]{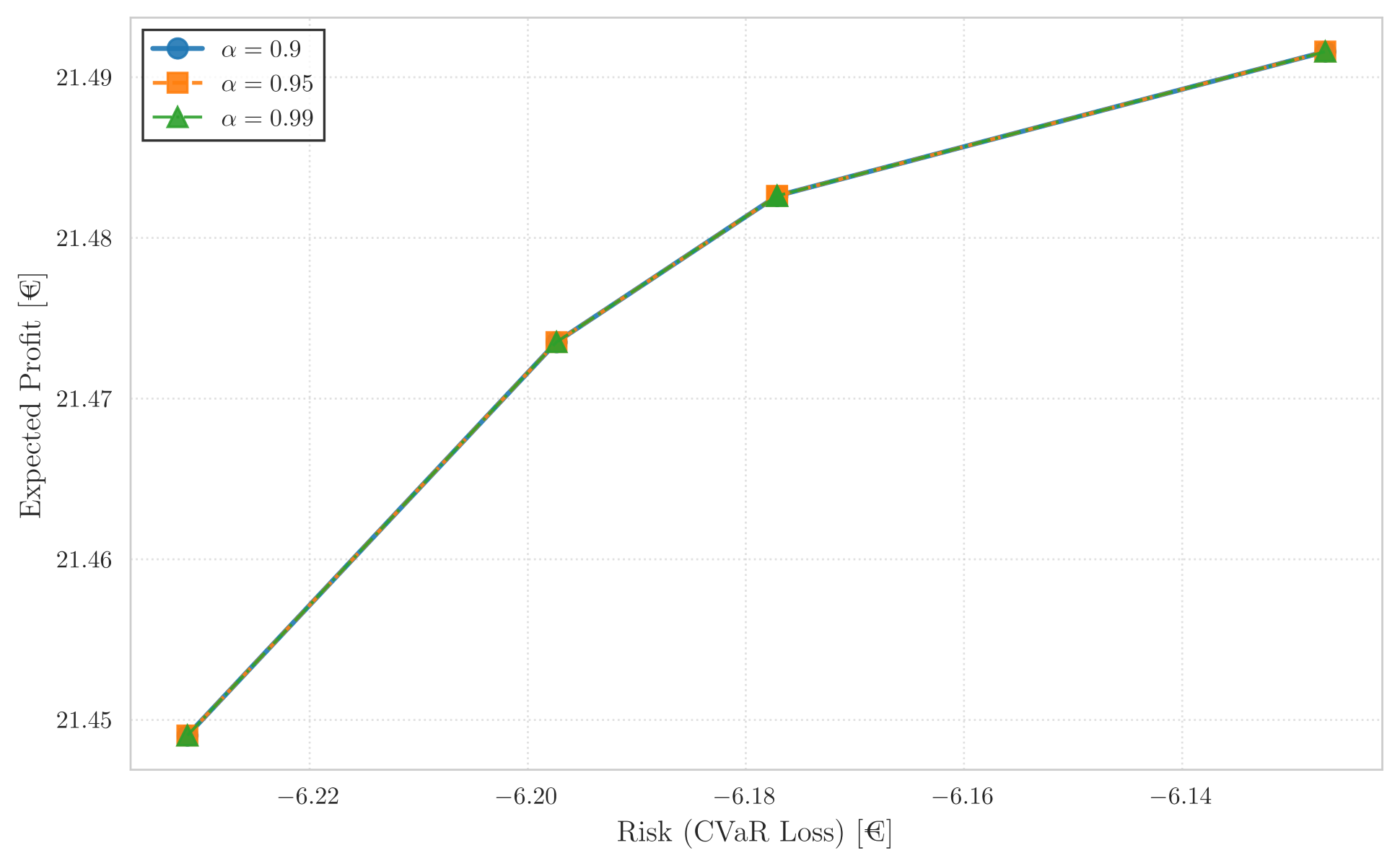}}
\caption{Risk-return trade-off in BESS multi-market participation.}
\label{fig}
\end{figure}

In Figure 4, the efficient frontiers of the Mean-CVaR optimization demonstrate the classic Pareto trade-off between maximizing expected profit and minimizing tail risk. Moving from right to left along the curve, the BESS operator can significantly reduce their risk exposure by sacrificing only a marginal amount of expected profit (dropping from roughly 21.49 € to 21.45 €). Furthermore, the near-perfect overlap of the curves across all tested confidence levels ($\alpha=0.90$, $0.95$, and $0.99$) highlights the structural robustness of the proposed framework. This convergence indicates that the optimal capacity allocation between the DA and mFRR markets remains remarkably stable, regardless of whether the operator is hedging against the worst 10$\%$ or the most extreme 1$\%$ of market realizations.

\begin{figure}[htbp]
\centerline{\includegraphics[width=0.5\textwidth]{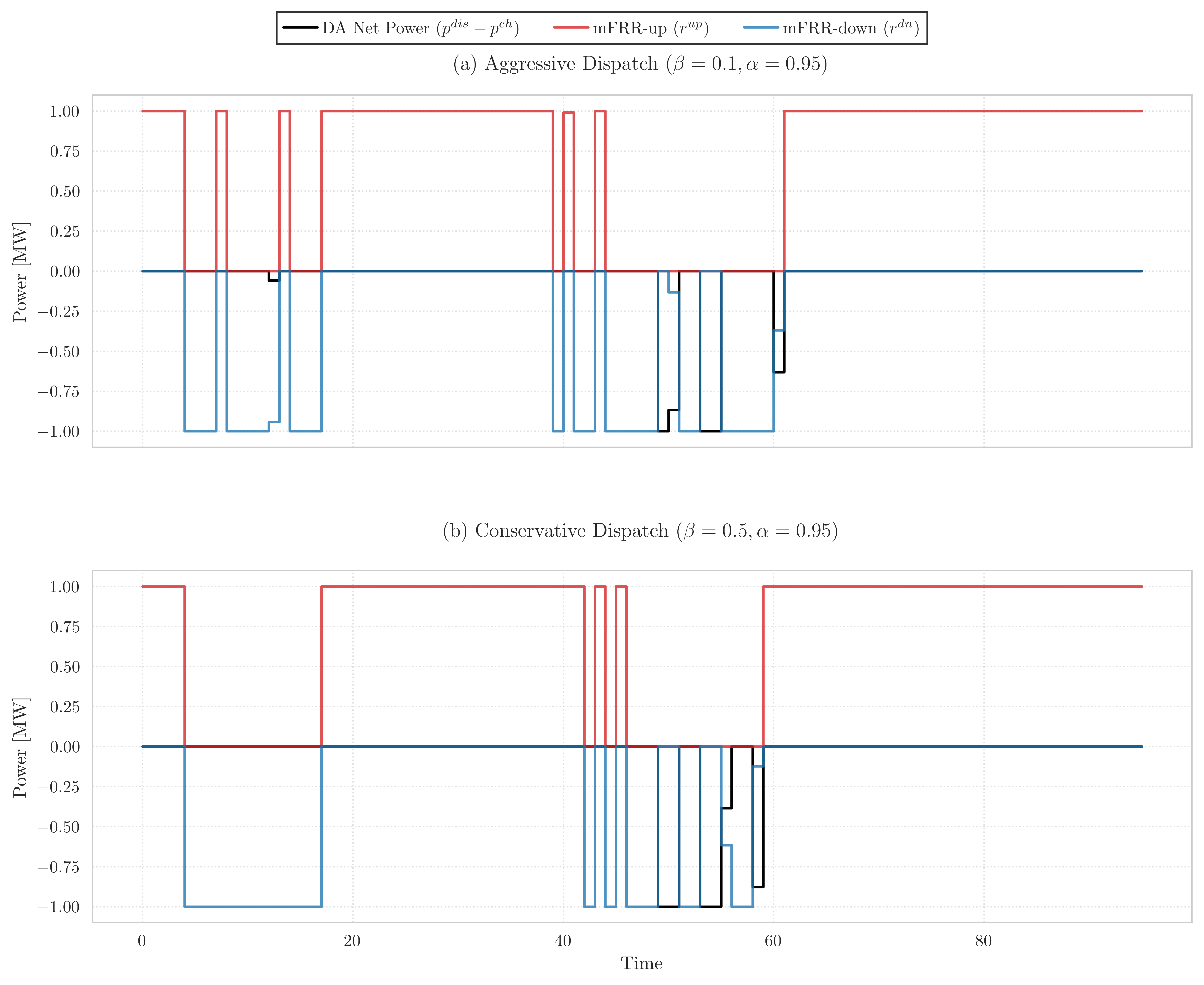}}
\caption{Optimized power dispatch in DA and mFRR markets.}
\label{fig}
\end{figure}

In Figure 5, the co-optimized power dispatch profiles for the DA and mFRR markets illustrate the operational shift induced by varying risk preferences. The figure captures the optimal charging and discharging behavior of the battery, demonstrating how the system balances revenue generation with physical risk mitigation. Subplot (a) shows the aggressive, profit-maximizing strategy ($\beta=0.1$), where the BESS frequently schedules intense DA charging and discharging cycles while concurrently allocating its maximum available capacity to mFRR reserves. While highly profitable under expected conditions, this rigid scheduling leaves the battery highly vulnerable to extreme energy activations. In contrast, subplot (b) demonstrates the conservative, risk-averse strategy ($\beta=0.5$)

\begin{figure}[htbp]
\centerline{\includegraphics[width=0.5\textwidth]{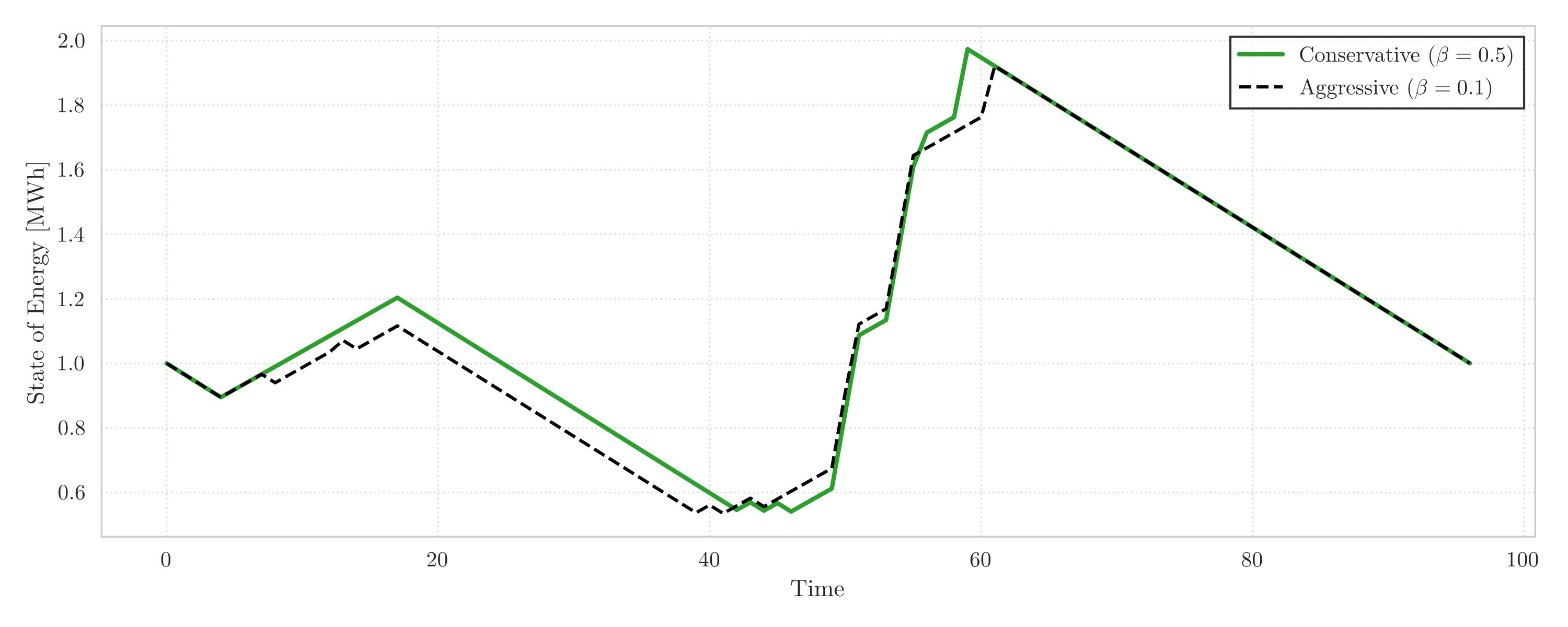}}
\caption{BESS energy level trajectories under varying risk preferences.}
\label{fig}
\end{figure}

In Figure 6, the evolution of the SoE is compared under aggressive ($\beta=0.1$) and conservative ($\beta=0.5$) bidding strategies. The plot clearly demonstrates that the risk-averse, conservative strategy maintains a higher energy buffer throughout most of the operating horizon (e.g., between $t=10$ and $t=40$). While the aggressive model exploits the battery more deeply to maximize expected DA and reserve revenues, the conservative model restricts these deeper discharges.

\section{Conclusion}
 
This paper presents a stochastic Mean-CVaR optimization framework for the co-optimization of BESS across the Day-Ahead and mFRR markets. By explicitly capturing the joint uncertainty of market prices and reserve activation through non-parametric scenario generation, the proposed approach addresses key limitations of conventional deterministic models. The integration of the CVaR metric enables a more realistic representation of financial risk, particularly under extreme market conditions, and provides a flexible mechanism to balance profitability and robustness.

The results highlight that modest levels of risk aversion can significantly improve revenue stability with only marginal reductions in expected profit. Moreover, the consistency of the efficient frontier across different confidence levels suggests that the model structure is inherently robust to varying degrees of uncertainty. From an operational perspective, the framework encourages more prudent energy management by maintaining sufficient flexibility to accommodate unexpected reserve activations. Overall, this work demonstrates the importance of incorporating both stochastic modeling and risk-aware decision-making in modern electricity markets, offering practical insights for BESS operators seeking to optimize multi-market participation.

\end{document}